\documentclass[aps,prl,nobalancelastpage,nobibnotes,twocolumn,superscriptaddress,10pt]{revtex4-2}

\usepackage[english]{babel}
\usepackage{color,amsthm,amsmath,amsxtra,amsfonts,dsfont,graphicx,bm,amssymb}
\usepackage[hidelinks]{hyperref}
\usepackage[utf8]{inputenc}
\usepackage{mathtools}
\usepackage{braket}
\usepackage{xcolor}
\usepackage[T1]{fontenc}
\usepackage{tabularx}
\usepackage{orcidlink}

\newcommand{\sect}[1]{{\bf #1.---} }

\newcommand{\figref}[1]{Fig.~\ref{#1}}

\newcommand{\eqeqref}[1]{Eq.~(\ref{#1})}

\newcommand{\tr}[1]{\operatorname{tr}\mathopen{}\left[#1\right]}

\begin{document}

    \title{Universality in Quantum Many-Body Spin Ratchets} 
    
    \author{Andrea Azzali\,\orcidlink{0009-0004-1786-1859}}
    \affiliation{Technical University of Munich, TUM School of Natural Sciences, Physics Department, James-Franck-Str. 1, 85748 Garching, Germany}
    \affiliation{Munich Center for Quantum Science and Technology (MCQST), Schellingstr. 4, München 80799, Germany}
    \author{Marko Ljubotina\,\orcidlink{0000-0003-0038-7068}}
    \email{marko.ljubotina@tum.de}
    \affiliation{Technical University of Munich, TUM School of Natural Sciences, Physics Department, James-Franck-Str. 1, 85748 Garching, Germany}
    \affiliation{Munich Center for Quantum Science and Technology (MCQST), Schellingstr. 4, München 80799, Germany}

    \begin{abstract}
        We consider a class of quantum circuits, known as quantum many-body spin ratchets, which act on alternating spin chains with two distinct spin types on even and odd sites. 
        This setup breaks parity symmetry and exhibits chiral transport.
        It has been shown that the drift velocity in the case of integrable, SU(2)-symmetric quantum spin ratchets is universal, and conjectured that the same holds in the non-integrable case.
        Here, we prove that this universal drift is exact at all times, requiring only translation invariance and U(1) symmetry.
        Furthermore, we prove that it is sufficient for the circuit to be translation-invariant on average by deriving the same result in the random case.
        Finally, we demonstrate that breaking translation invariance immediately results in deviations from this universal prediction.
    \end{abstract}
    
    \maketitle

    \sect{Introduction}
    Determining the dynamics of quantum many-body systems is a fundamental problem in modern theoretical physics, but it poses several theoretical and computational challenges: perturbative methods often fail to capture the strongly interacting regime, and numerical simulations must handle Hilbert spaces whose dimensions grow exponentially with system size. 
    In this context, many efforts aim to find non-perturbative, universal descriptions that depend not on the microscopic details of the models but on general features, such as symmetries.

    In this work, we focus on dynamics in systems lacking parity symmetry, and more specifically on the presence of chiral transport. 
    The dynamical properties of systems with broken reflection symmetry have been a long-standing topic of interest in theoretical physics, ranging from the well-known classical examples of exclusion processes~\cite{derrida_exact_1993,corwin_kardarparisizhang_2012,essler_lifted_2025} to more recent works on quantum systems~\cite{orito_entanglement_2023,li_observation_2024,pang_synthetic_2024}.
    
    With the recent advances in digital quantum devices, a particularly relevant framework for investigating transport in quantum many-body systems is provided by quantum circuits, which can describe a wide variety of phenomena, such as the spreading of correlations and the growth of entanglement~\cite{gong_coarse-grained_2022}, and also allow for hydrodynamic descriptions~\cite{khemani_operator_2018,rakovszky_diffusive_2018,hubner_generalized_2025,p_kim_circuits_2026}.
    Not only can such systems be studied in experimental setups based on ultracold atoms, trapped ions, superconducting qubits, and others~\cite{noel_measurement-induced_2022,daley_practical_2022,bravyi_future_2022,ma_high-fidelity_2023,google_quantum_ai_and_collaborators_measurement-induced_2023,haghshenas_digital_2026}, 
    but they can also admit analytical solutions~\cite{vanicat_integrable_2018,bertini_exact_2019,borsi_construction_2022,znidaric_integrability_2025,paletta_integrability_2025,bertini_exactly_2026}, which can then be tested against either numerical or experimental results. 

    In order to address transport in parity-broken systems, we will focus on a class of circuits known as \textit{quantum many-body spin ratchets}~\cite{Zadnik2024}. 
    This choice is motivated by the minimal setup needed to capture general behavior while neglecting model-specific details.
    It also shares the main features of asymmetry, periodicity, and Floquet driving present in several other one-dimensional systems in which chiral transport has already been studied~\cite{reimann_quantum_1997,denisov_periodically_2007,hanggi_artificial_2009,salger_directed_2009,hamamoto_scaling_2019}. 
    
    Our setup will thus consist of an alternating spin chain whose unit cell comprises two sites with different spins $s_1 \neq s_2$. 
    A circuit then acts on this chain in a brickwork pattern, such that each spin-$s_1$ ($s_2$) propagates to the right (left). 
    From a quantum information viewpoint, the topological index for these ratchets is nonzero, a hallmark of chiral information propagation~\cite{gross_index_2012,po_chiral_2016,cirac_matrix_2017,gong_topological_2021,gong_coarse-grained_2022}. 
    In this work, however, we will address charge transport in the linear-response regime, characterized by the spin-spin correlation function.
    So far, this two-point function was observed to display a drift in the case of integrable, SU(2)-symmetric circuits~\cite{Zadnik2024}. 
    In that example, integrability was used to derive an analytical expression for the drift velocity, which showed surprising universality: it is independent of the exact form of the gates used to define the circuit. 
    Furthermore, it was shown through numerical simulations that the same drift velocity appears in non-integrable SU(2)-symmetric quantum many-body spin ratchets.
    
    In the present work, we prove that in generic U(1)-conserving quantum many-body spin ratchets the two-point function drifts with this universal velocity exactly, at all times, and that the velocity depends only on the sizes of the two spins in the chain and on the background magnetization.
    We show that the only property required for these universal results to hold is the translation invariance of the circuit. 
    These assumptions are very general and encompass the two prototypical cases studied in quantum circuits: the exactly homogeneous circuit, in which the same gate is applied at every step, and the random circuit. 
    The former allows direct control over the dynamics in terms of the single-gate parameters, while the latter often benefits from numerous simplifications, making analytic treatment easier~\cite{nahum_quantum_2017,nahum_operator_2018,zhou_emergent_2019,fisher_random_2023}.
    Moreover, we note that our derivation can be viewed as a circuit realization of an exact first-moment sum rule known from hydrodynamics~\cite{spohn_nonlinear_2014,doyon_lecture_2020,doyon_free_2021}, and the universality has a simple origin: in a ratchet, the average current carried by the Gibbs state is fixed by the swap operations alone, independently of the gates.

    Finally, we verify that charge conservation and translation invariance are the necessary requirements for the universal drift velocity. 
    This follows intuitively from the definition of the drift velocity, which is obtained from the two-point function of the magnetization along $z$ by taking its first moment. 
    We confirm this intuition by numerically studying the dynamics in systems that are not invariant under translation by a single unit cell and observing the breakdown of the universal result.

\begin{figure}
        \centering
        \includegraphics[width=\linewidth]{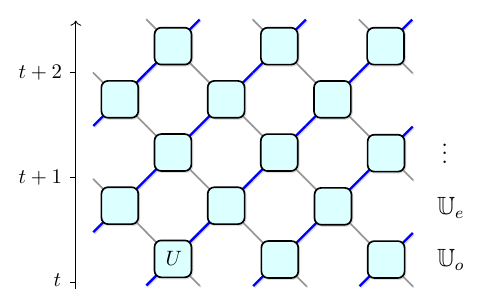}
        \caption{Graphical depiction of the ratchet circuit. Here, only 
        five layers are shown. The legs of the tensors acting on $\mathcal H_{s_{1,2}}$ have been colored in blue and gray, respectively.}
        \label{fig:setup}
    \end{figure}

    \sect{Model}
    We consider quantum many-body spin ratchets, introduced in~\cite{Zadnik2024}, and briefly review their construction here. A quantum many-body spin ratchet is a quantum circuit acting on a spin chain in which the odd sites are spin-$s_1$ and the even sites are spin-$s_2$. We denote the local Hilbert spaces by $\mathcal H_{s_1}$ and $\mathcal H_{s_2}$ respectively. The unitary evolution $\mathbb U$ corresponding to a timestep consists of two layers, $\mathbb U_e$ and $\mathbb U_o$, which are products of gates in a brickwork fashion (see \figref{fig:setup}): 
    \begin{equation}
        \mathbb U_o = \bigotimes_{j} U_{2j+1,2j+2}, \qquad \mathbb U_e = \bigotimes_{j} U_{2j,2j+1},
    \end{equation}
    where the subscript denotes the sites on which the gate acts.
    The gates $U$ are defined on the following spaces:
    \begin{equation*}
        U_{2j+1,2j+2}: \mathcal H_{s_1} \otimes \mathcal H_{s_2} \rightarrow \mathcal H_{s_2} \otimes \mathcal H_{s_1}
    \end{equation*} and analogously for the other layer. 
    They are the composition of a unitary $V$ and a ``swap'' $P$, which acts as
    \begin{equation}
        P ~ \ket{\psi_1} \otimes \ket{\psi_2} = \ket{\psi_2} \otimes \ket{\psi_1}
    \end{equation}
    on $\ket{\psi_{1,2}} \in \mathcal H_{s_{1,2}}$.
    
    This setup, which breaks parity symmetry, allows us to describe an inhomogeneous chain with different numbers of degrees of freedom propagating to the right and left.
    We are interested in the dynamics with global conservation of the magnetization along $z$,
    \begin{equation}
        S^z_\text{tot} = \sum_j Z_{2j+1,2j+2}, ~ ~ ~ Z_{2j+1,2j+2} = S^z_{2j+1} + S^z_{2j+2} .
    \end{equation}
    Following~\cite{Zadnik2024}, we study the ``drift'' of the connected two-point function $S(j,t) = \langle Z_{1,2}(0) ~ Z_{2j+1,2j+2} (t) \rangle_c$ by computing, at each time, its center of mass
    \begin{equation}
        \bar x(t) = \frac{\sum_j j S(j,t)}{\sum_j S(j,t)} .
        \label{eq:center-of-mass}
    \end{equation}
    We consider as initial states the most generic Gibbs states
    \begin{equation}
        \rho_\mu = \frac{\exp(-\mu S^z_\text{tot})}{\tr{\exp(-\mu S^z_\text{tot})}} ,
    \end{equation}
    since the total magnetization along $z$ is the only conserved charge, except in special cases.
    It was shown in Ref.~\cite{Zadnik2024}, using generalized hydrodynamics, that for integrable ratchets with SU(2)-symmetric gates the two-point function drifts with velocity $v_d$, given by
    \begin{equation}
        v_d(\mu) = \frac{\partial_\mu \langle S^z_1 \rangle - \partial_\mu \langle S^z_2 \rangle}{\partial_\mu \langle S^z_1 \rangle + \partial_\mu \langle S^z_2 \rangle}
        \label{eq:drift-velocity}
    \end{equation}
    with $\langle \cdot \rangle$ denoting the expectation value $\tr{(\cdot) \rho_\mu}$.
    In the $\mu = 0$ case, \eqeqref{eq:drift-velocity} simplifies to
    \begin{equation}
        v_d(\mu=0) = \frac{s_1(s_1+1) - s_2(s_2+1)}{s_1(s_1+1) + s_2(s_2+1)},
        \label{eq:drift-velocity-mu-0}
    \end{equation}
    which is just a function of the Casimir operator eigenvalues for the spins $s_1$ and $s_2$.
    It was also conjectured~\cite{Zadnik2024} that this should hold for non-integrable ratchets with SU(2)-symmetric gates. In the following sections, we prove that, in fact, $\bar x(t) = v_d t$ holds exactly at all times in any U(1)-conserving translation-invariant quantum spin ratchet. 

    \sect{Translation-invariant U(1)-symmetric circuits}
    Let us assume that each unitary applied is a fixed gate conserving the two-site charge: $[Z_{\ell,\ell+1}, V_{\ell,\ell+1}] = 0$. To prove that the drift velocity is indeed constant and given by \eqeqref{eq:drift-velocity}, we determine $\bar x(t)$ explicitly.
    The denominator of \eqeqref{eq:center-of-mass} is constant and given by $-(\partial_\mu \langle S^z_1 \rangle + \partial_\mu \langle S^z_2 \rangle)$. %
    \footnote{$\partial_\mu \langle S^z_i \rangle = - \langle (S^z_i)^2 \rangle + \langle S^z_i \rangle^2$ for $i = 1,2$.}
    To study the numerator, we introduce the staggered magnetization
    \begin{equation}
        \mathcal M = \sum_j (S^z_{2j+1} - S^z_{2j+2}).
    \end{equation}
    
\begin{figure}
        \centering
        \includegraphics[width=\linewidth]{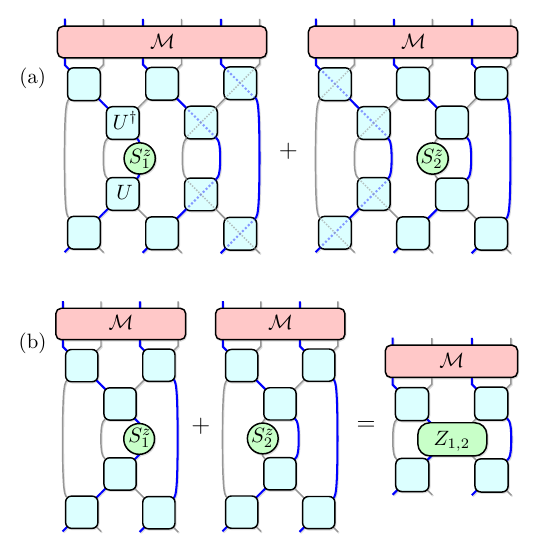}
        \caption{Sketch of the proof that $\langle Z_{1,2}(0) \mathcal M(t) \rangle_c$ is constant. For clarity, we only show the $\mu=0$ case and two brickwork layers.
        This expectation value can be rewritten as in \eqeqref{eq:staggered-sum-12}, and unitarity allows for simplifications denoted by dashed lines in (a). The resulting tensor network, because of translation invariance and charge conservation, is then equivalent to the initial one, but with only one layer of evolution (b). }
        \label{fig:proof-transl-inv}
    \end{figure}

    We then make an important remark, namely that $\langle Z_{1,2}(0) ~ \mathcal M(t) \rangle_c = -(\partial_\mu \langle S^z_1 \rangle - \partial_\mu \langle S^z_2 \rangle)$ is conserved.
    To prove this statement, we rewrite the expression as
    \begin{equation}
        \label{eq:staggered-sum-12}
        \langle Z_{1,2}(0) ~ \mathcal M(t) \rangle_c = \langle S^z_1(0) ~ \mathcal M(t) \rangle_c + \langle S^z_2(0) ~ \mathcal M(t) \rangle_c
    \end{equation}
    (see \figref{fig:proof-transl-inv}a). By unitarity, this expression has the same expectation value as the l.h.s.\ of \figref{fig:proof-transl-inv}b. The identity in \figref{fig:proof-transl-inv}b then follows by translation invariance and charge conservation. Finally, we notice that the r.h.s.\ is the initial expression for $\langle Z_{1,2}(0) ~ \mathcal M(t) \rangle_c$ from which one layer has been removed. Hence, we can conclude that this quantity is constant. \footnote{While \figref{fig:proof-transl-inv} only shows the $\mu = 0$ case, a generalization to $\mu \neq 0$ is possible: in the tensor networks of \figref{fig:proof-transl-inv}, one has to insert $\rho_\mu$, which is no longer proportional to the identity. Using the fact that $\rho_\mu$ commutes with time evolution, it can be put on top of the tensor network, multiplying $\mathcal M$.
    Given the simple structure of $\rho_\mu \propto \bigotimes_j \exp \left (-\mu Z_{2j+1,2j+2} \right )$, the translation invariance argument used above can be replicated.}
    
    We now proceed to determine the increase in the numerator of \eqeqref{eq:center-of-mass} over a timestep, and begin by addressing the contributions of the ``swap'' operations $P$, which evolve $\sum_j j \left ( S^z_{2j+1} + S^z_{2j+2} \right )$ into \begin{align}
        & \sum_j j \left ( S^z_{2(j-1)+1} + S^z_{2(j+1)+2} \right ) \nonumber\\ = & \sum_j (j+1) S^z_{2j+1} + \sum_j (j-1) S^z_{2j+2} \\ = & \sum_j j \left ( S^z_{2j+1} + S^z_{2j+2} \right ) + \sum_j \left ( S^z_{2j+1} - S^z_{2j+2} \right ) .\nonumber
    \end{align}
    Hence, the action of the $P$ operators amounts to adding a term with the staggered magnetization $\langle Z_{1,2}(0) \mathcal M(t) \rangle_c$ to the numerator of \eqeqref{eq:center-of-mass}. To determine, instead, how the unitaries act on that numerator, we first observe that the unitaries $V_{2j+1,2j+2}$ commute with $Z_{2j+1,2j+2}$ and thus play no role. To deal now with the unitaries $V_{2j,2j+1}$, we rewrite $\sum_j j \left ( S^z_{2j+1} + S^z_{2j+2} \right )$ as
    \begin{equation}
    \begin{split}
        & \sum_j \left ((j-1) S^z_{2j} + j S^z_{2j+1} \right ) \\ = & \sum_j \left ( j \!-\! \frac 1 2 \right )\! \left ( S^z_{2j} + S^z_{2j+1} \right )\! +\! \frac 1 2 \sum_j \left ( S^z_{2j+1} - S^z_{2j} \right ).
        \end{split}
    \end{equation}
    This comprises a term with $Z_{2j,2j+1}$, which is left untouched, and another term with the staggered magnetization, which is unaffected as shown above.
    Hence, the increase in the numerator of \eqeqref{eq:center-of-mass} over a timestep is entirely due to the $P$ gates, and equals $\langle Z_{1,2}(0) \mathcal M(t) \rangle_c$.
    Substituting in \eqeqref{eq:center-of-mass}, we obtain
    \begin{equation}
    \label{eq:increase-cm}
        \bar x(t+1) - \bar x(t) = \frac{\langle Z_{1,2}(0) \mathcal M(t) \rangle_c}{\sum_j S(j,t)}
    \end{equation}
    and find the analytical expression $\bar x(t) = v_d t$, with $v_d$ given by \eqeqref{eq:drift-velocity}.

    This derivation is a circuit realization of an exact first-moment sum rule familiar from hydrodynamics~\cite{spohn_nonlinear_2014,doyon_lecture_2020,doyon_free_2021}.
    In a translation-invariant system with a local conservation law, the change of $\sum_j j S(j,t)$ over a time step is the correlation of $Z_{1,2}(0)$ with the total current $J(t)$, i.e., the total charge transferred across the boundaries between unit cells during that step.
    Translation invariance gives $\langle Z_{1,2}(0) J(t) \rangle_c = L^{-1} \langle S^z_\text{tot} J(t) \rangle_c$, with $L$ the number of unit cells, and charge conservation together with the stationarity of $\rho_\mu$ make the latter time-independent; \figref{fig:proof-transl-inv} is a diagrammatic version of this step.
    In our setting, $\mathcal M$ is the contribution of the swaps to $J$, while the gates $V$ do not contribute to the correlation.
    Since $\partial_\mu \langle \mathcal O \rangle = - \langle S^z_\text{tot} \mathcal O \rangle_c$, \eqeqref{eq:drift-velocity} can be read as $v_d = \partial_\mu \bar\jmath / \partial_\mu \bar q$, the Euler-scale velocity of a single conserved charge, where $\bar q = \langle S^z_1 \rangle + \langle S^z_2 \rangle$ is the mean charge per unit cell and $\bar\jmath = \langle S^z_1 \rangle - \langle S^z_2 \rangle$ the mean charge transferred between neighboring unit cells per time step.
    The universality of $v_d$ thus stems from the fact that, in a quantum many-body spin ratchet, the mean current is fixed by the swaps alone: since each $V$ commutes with the local factors of $\rho_\mu$, the gates transport no charge on average.
    We also note that the argument does not require the same gate to be applied at every time step; it suffices that each layer is translation-invariant.
    
\begin{figure*}
            \centering
            \includegraphics[width=0.85\linewidth]{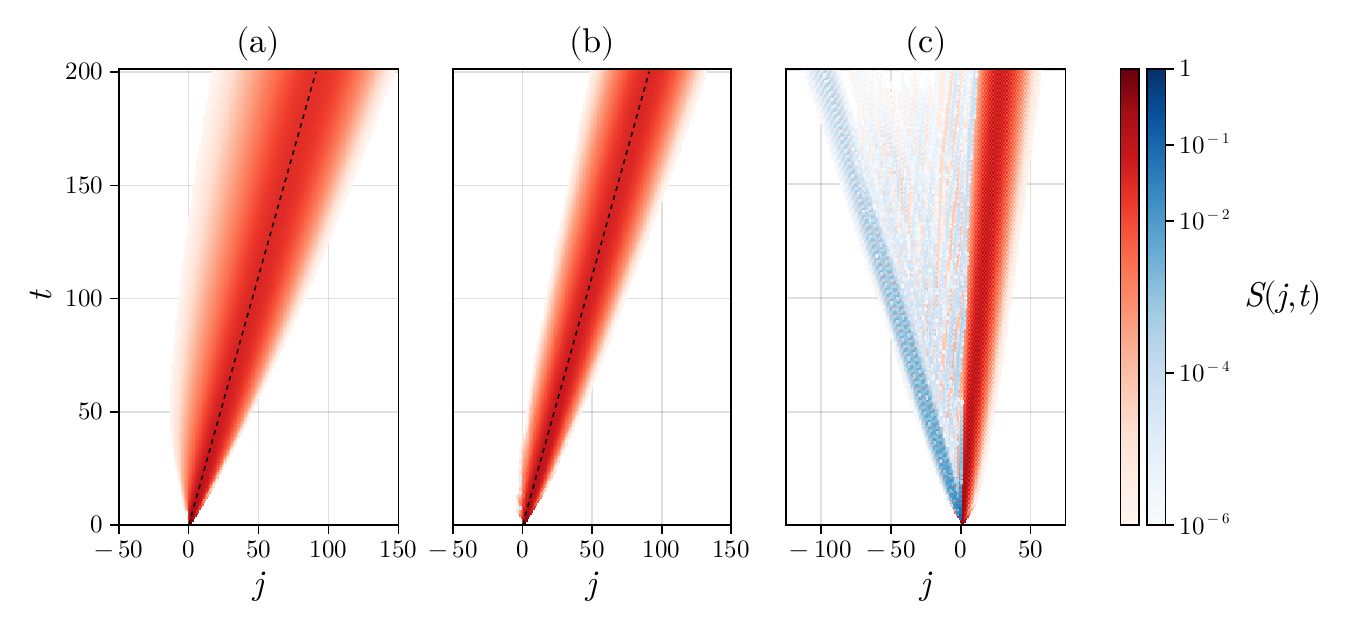}
            \caption{Two-point function $S(j,t)$ at $\mu=0$ for three different setups, with $s_1 = 1$, $s_2 = \frac 1 2$. The heatmaps show positive (negative) values in red (blue); all plots have the same scales on both axes. Panel (a): Homogeneous circuit, in which all gates are equal to a fixed, randomly chosen unitary $A$ conserving charge (see Supplemental Material); the drift velocity is consistent with the analytical prediction $v_d = 5/11$ (dashed line). Panel (b): Each charge-conserving gate is independently sampled from the Haar distribution, without averaging. After the first few iterations, the drift velocity again matches the analytics (dashed line). Panel (c): the alternating pattern between $G$ and $\mathbb I$, breaking translation invariance and exhibiting a different $v_d \approx 0.14$.}
            \label{fig:plots-haar-break}
    \end{figure*}

    To provide intuition, we note that the presence of the staggered magnetization $\mathcal M$ in \eqeqref{eq:increase-cm} is due to the fact that the spin-$s_1$ and the spin-$s_2$ degrees of freedom travel in two opposite directions, thus contributing with opposite signs to the velocity of the center of mass $\bar x(t)$ (see \figref{fig:plots-haar-break}a for numerical results).

    \sect{Random U(1)-symmetric circuits}
    We now address the case of random charge-conserving gates, drawn from a probability distribution.
    As long as each gate conserves the total magnetization along $z$, it is possible to relax the translation invariance assumption, only demanding that translation invariance be satisfied after averaging over the probability distribution.
    
    We can prove this statement by replicating the reasoning of the previous section. The denominator of \eqeqref{eq:center-of-mass} is always exactly conserved and given by $-(\partial_\mu \langle S^z_1 \rangle + \partial_\mu \langle S^z_2 \rangle)$. However, the conservation of $\langle Z_{1,2}(0) \mathcal M(t) \rangle_c$ will now only hold on average. The proof can be reproduced by replacing the unitary evolution of operators $U^\dagger (\cdot) U$ by its average $\int \mathrm{d} U ~ U^\dagger (\cdot) U$, and noticing that this map is still translation-invariant and leaves the identity and the total magnetization unaffected. We also observe that the exactly translation-invariant setup of the previous section can be viewed as a particular instance of this problem, in which the probability distribution is delta-like.
    
    Numerically, we simulated the averaged evolution $\int \mathrm{d} U ~ U^\dagger (\cdot) U$ with respect to the Haar measure (see End matter) and observed the correct drift velocity $v_d$. We also considered inhomogeneous circuits in which each charge-conserving unitary $V$ was sampled from the Haar measure (see \figref{fig:plots-haar-break}b). In this case, lacking translation invariance, the universal value of the drift velocity of \eqeqref{eq:drift-velocity} is not reached immediately, but after a short, non-universal timescale (at $t \sim 10$ for $s_1 = 1$, $s_2 = \frac 1 2$), a common behavior in random circuits.

    \sect{Breaking translation invariance}
    We now address the necessity of the translation invariance hypothesis, describing violations of \eqeqref{eq:drift-velocity} for the drift velocity when this assumption is weakened. To show this, we provide counterexamples in the simplest non-trivial setting, namely at $\mu=0$, $s_1 = 1$, $s_2 = \frac 1 2$, and choosing each unitary $V$ to be either the identity $\mathbb I$ or the unitary $G$ that moves a charge $1$ between the sites (when possible). More precisely, $G$ is defined as follows: on the $Z_{1,2}=\pm \frac 3 2$ subspaces, it acts as the identity; elsewhere, it satisfies
    \begin{equation*}
        G \ket{\pm 1} \otimes \ket{\mp\frac 1 2} = \ket{0} \otimes \ket{\pm \frac 1 2},
    \end{equation*}
    \begin{equation*}
        G \ket{0} \otimes \ket{\pm \frac 1 2} = \ket{\pm 1} \otimes \ket{\mp\frac 1 2}.
    \end{equation*}
    Since $G^\dagger Z_{1,2} G$ is still diagonal in the $S^z$ basis, it is straightforward to study the evolution of the operator $Z_{1,2}$ in such ratchets, and verify that \eqeqref{eq:drift-velocity-mu-0} may be violated.
    
    We also consider circuits in which $\mathbb I$ and $G$ are alternated within each layer, or similar patterns in which $G$ is applied once every $m$ gates in each layer. In these setups, whose period is larger than the unit cell of the system, we numerically observe drift velocities different from the ones predicted by \eqeqref{eq:drift-velocity-mu-0}, suggesting that translation invariance is indeed a necessary condition in order to obtain the universal drift velocity (see \figref{fig:plots-haar-break}c for $m=2$).

    We note that it is possible to block gates together so that the resulting circuit, with gates acting on a larger space, is homogeneous (for the alternating pattern, with $m = 2$, each block would contain $4$ gates and act on a space of dimension $(2s_1+1)^2 \cdot (2s_2+1)^2$).
    These new gates, however, do not commute with the two-site magnetization $Z_{1,2}$, but only with its four-site analog $\mathcal Z_{4n} = \sum_{\ell = 0}^3 S^z_{4n+\ell}$.
    Considering this coarse-grained operator, the circuit is again U(1)-symmetric and homogeneous after the blocking; hence, for the two-point function
    \begin{equation}
        \label{eq:expanded-4sites-Z}
        \left \langle \mathcal Z_{0} (0) ~ \mathcal Z_{4n} (t) \right \rangle_c = \langle Z_{1,2} (0) ~ Z_{4n+1, 4n+2} (t) \rangle_c + \ldots
    \end{equation}
    the drift velocity is still given by \eqeqref{eq:drift-velocity-mu-0}. However, this does not fix the drift velocity for the various terms on the r.h.s.\ of \eqeqref{eq:expanded-4sites-Z}, including the two-point function for the operator $Z$.
    
    \sect{Conclusions and Outlook}
    Within the setup of quantum many-body spin ratchets, we have studied chiral transport arising from the explicit breaking of parity. In particular, we have shown that the universal drift velocity of the two-point function found in Ref.~\cite{Zadnik2024} is exact at all times, requiring only charge conservation and translation invariance on average, and we have identified our derivation with the first-moment sum rule of hydrodynamics. It should be noted that the drift is only due to the swap operations, which enable chiral information transport with a nonzero topological index. Indeed, choosing all gates $V$ to be the identity still yields the drift velocity of \eqeqref{eq:drift-velocity}~\cite{Zadnik2024}. In agreement with this, the absence of drift was previously found in integrable parity-broken circuits without the swaps $P$~\cite{paletta_integrability_2025}. 
    
    We also observe that the argument used to prove that $\langle Z_{1,2}(0) \mathcal M(t) \rangle_c$ is constant does not rely on $\mathcal M$ or on the specific choice of conserved charge $Z$, as expected from its equivalence with the sum rule, allowing for further generalizations. However, it does not seem possible to relax the symmetry requirements: concerning translation invariance, non-homogeneous circuits display different drift velocities for a given fixed starting cell, even when some periodicity is preserved in the setup. Concerning charge conservation, random unitary gates without any symmetry also violate \eqeqref{eq:drift-velocity}. 
    
    It remains to investigate whether symmetry arguments could constrain higher moments of the two-point function, for instance, to determine whether the fluctuations around $\bar x$ are symmetric or skewed. A similar question arises for the first moments of current fluctuations, which have been determined in integrable ratchets using Generalized Hydrodynamics~\cite{Zadnik2024}. In this case, the currents exhibit an asymmetry due to the breaking of time-reversal symmetry, indicating that the Gibbs state is not in thermodynamic equilibrium. However, this statement should also be verified in the generic case.

    A natural extension concerns ratchets with larger spins, in which the gates may conserve several charges $Q^a = \sum_j q^a_j$, with local densities $q^a_j$ defined on the unit cells, while still generating nontrivial dynamics.
    The sum rule holds for any such charge and fixes the first moments of all two-point functions $S^{ab}(j,t) = \langle q^a_0(0) ~ q^b_j(t) \rangle_c$ in the generalized Gibbs states $\rho_{\bm\mu} \propto \exp(-\sum_a \mu_a Q^a)$, namely $\sum_j S^{ab}(j,t) = C^{ab}$ and $\sum_j j S^{ab}(j,t) = B^{ab} t$, with $C^{ab} = -\partial_{\mu_a} \bar q^b$ and $B^{ab} = -\partial_{\mu_a} \bar\jmath^b$ given by the mean density and current of charge $b$, defined as for the magnetization.
    When the densities are sums of single-site terms, $\bar\jmath^b$ is fixed by the swaps alone, so that each charge drifts with its own universal velocity $v_d^a = B^{aa}/C^{aa}$.
    However, the charges are coupled whenever $B$ and $C$ have nonzero off-diagonal elements, and the velocities of the normal modes at the Euler scale are expected to be given by the eigenvalues of the flux Jacobian $B C^{-1}$~\cite{spohn_nonlinear_2014,doyon_lecture_2020}, which generally differ from the $v_d^a$.
    A charge with $v_d^a = 0$ could therefore still be transported chirally, with $S^{aa}(j,t)$ splitting into components moving in opposite directions while its center of mass remains at rest.
    Whether such hidden chiral transport arises, and how it manifests beyond the first moment, remains an interesting open question.

    \sect{Acknowledgments} The authors acknowledge useful discussions with Friedrich H\"ubner and Bruno Bertini.
    The authors acknowledge support by the Deutsche Forschungsgemeinschaft (DFG, German Research Foundation) under Germany’s Excellence Strategy – EXC-2111 – 390814868. 
    M.L. acknowledges funding from the European Union's Horizon Europe research and innovation programme under the Marie Sk\l{}odowska-Curie grant agreement No.~101279178 (PSymQDyn).
    Funded by the European Union. Views and opinions expressed are, however, those of the author(s) only and do not necessarily reflect those of the European Union or the European Research Executive Agency. Neither the European Union nor the granting authority can be held responsible for them.
    The tensor network simulations were performed using the ITensor library~\cite{fishman_itensor_2022}. The authors used Claude Opus 5.5 (Anthropic) to assist with editing the manuscript and drafting parts of the text; all content was directed, verified, and revised by the authors, who take full responsibility for it.

    \sect{Data availability} Numerical simulation codes and data are available on Zenodo~\cite{azzali_universality_2026}.

    \bibliography{bib}

    \section{End matter}
    \sect{Random U(1)-symmetric circuits}
    As an example, we choose a specific distribution -- namely, the Haar distribution -- for random charge-conserving gates, and discuss their action on operators by analytically computing the average~\cite{rakovszky_sub-ballistic_2019,fisher_random_2023}.
    
    To do so, we consider a unitary $V$ conserving the magnetization along $z$ and write it in block-diagonal form, taking each block, labeled by total magnetization as $V^{-(s_1 + s_2)}, \ldots, V^{s_1 + s_2}$, to be independently Haar-distributed.
    The averaged time evolution induced by these gates can then be studied by determining the entries of the tensor
    \begin{equation}
        \int \mathrm{d} V ~ V_{(i_1,i_2),(j_1,j_2)} V^*_{(i'_1,i'_2),(j'_1,j'_2)},
        \label{eq:averaged-evolution}
    \end{equation}
    where the indices label the magnetization,
    i.e. $i_{1,2} \in \{-s_{1,2}, \ldots , +s_{1,2}\}$, and the same for $i',j,j'$. In this notation,
    \begin{equation}
        V_{(i_1,i_2),(j_1,j_2)} = \bra{i_1 i_2} V \ket{j_1 j_2} .
    \end{equation}
    The only nonzero entries of the linear operator \eqref{eq:averaged-evolution} are those in which
    \begin{itemize}
        \item $i_1 + i_2 = j_1 + j_2$ and $i'_1 + i'_2 = j'_1 + j'_2$ by charge conservation;
        \item $i_1 + i_2 = i'_1 + i'_2$: if, instead, the indices of $V$ and $V^*$ belong to different sectors, then the average factorizes since different sectors are independent, and $\int \mathrm{d} V ~ V = 0$ implies that the entry is zero.
    \end{itemize}
    Hence, all indices must lie in the same sector $Z_{1,2} = m$ for the entries to be nonzero. In this case, their value is obtained by Weingarten calculus:
    \begin{equation*}
        \int \mathrm{d} V ~ V^{m}_{(i_1,i_2),(j_1,j_2)} (V^{m})^*_{(i'_1,i'_2),(j'_1,j'_2)} = \frac{1}{d_m} \prod_{\alpha=1,2}\delta_{i_\alpha,i'_\alpha} \delta_{j_\alpha,j'_\alpha},
    \end{equation*}
    where $d_m$ is the dimension of the sector. Now, an operator 
    $\mathcal O$ will be mapped into another operator $\mathcal O' = \int \mathrm{d} V ~ V^\dagger \mathcal O V$ whose entries are
    \begin{equation*}
        \mathcal O'_{(j_1,j_2),(j'_1,j'_2)} = \int \mathrm{d} V ~ V^*_{(i_1,i_2),(j_1,j_2)} \mathcal O_{(i_1,i_2),(i'_1,i'_2)} V_{(i'_1,i'_2),(j'_1,j'_2)},
    \end{equation*}
    where we leave the summation over $i$,$i'$ implicit.
    This operator $\mathcal O'$ is diagonal; it only depends on the diagonal entries of $\mathcal O$, and its entries are
    \begin{equation*}
        \mathcal O'_{(j_1,j_2),(j_1,j_2)} = \frac{1}{d_{j_1+j_2}} \sum_{i_1+i_2=j_1+j_2} \mathcal O_{(i_1,i_2),(i_1,i_2)},
    \end{equation*}
    i.e., the average over the diagonal entries within each sector. Finally, we note that averaging over random unitaries -- without the constraint of charge conservation -- yields $\int \mathrm{d}V ~ V^\dagger Z V = 0$, hence \eqeqref{eq:drift-velocity} no longer holds in this case.

    \clearpage
    \onecolumngrid

    \setcounter{section}{0}
    \setcounter{equation}{0}
    \setcounter{figure}{0}
    \setcounter{table}{0}
    \setcounter{page}{1}
    \renewcommand{\theequation}{S\arabic{equation}}
    \renewcommand{\thefigure}{S\arabic{figure}}
    \renewcommand{\thesection}{S\arabic{section}}
    \renewcommand{\thesection}{S\arabic{section}}

    \begin{center}
        \Large \textbf{Supplemental material for Universality in Quantum Many-Body Spin Ratchets}
    \end{center}
    \subsection*{Details and convergence of the numerical simulations}
    In this Section, we provide details on the numerical simulations. Throughout this work, we choose values of $s_{1,2}$ to be $s_1=1$, $s_2 = \frac 1 2$, which represent the smallest non-trivial case, and are therefore easiest to simulate numerically.
    
    The simulations evolve the operator $Z_{1,2}$, represented as a Matrix Product Operator (MPO), and then compute its expectation values. The operator $Z_{1,2}(t)$ is initially supported on only two sites and, at every step, on both sides of the MPO we add one site on which the operator acts trivially. In this way, we assume that we do not reach the boundary of the physical system during the simulation time. To perform the swap on a pair of neighboring sites, we contract the respective tensors in the MPO and then perform a singular value decomposition, demanding that the output tensors have the site indices swapped. 
    
\begin{figure}[h]
        \centering
        {\includegraphics[width=0.49\linewidth]{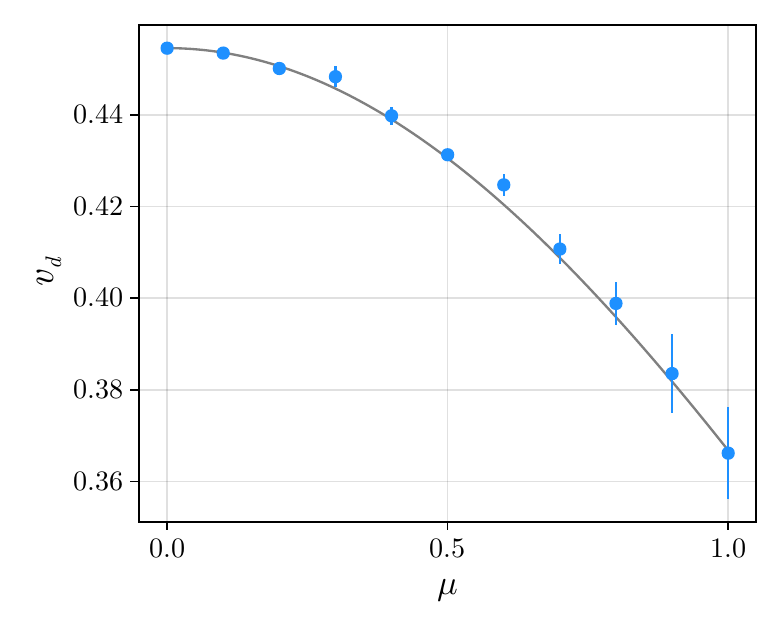}}
        {\includegraphics[width=0.49\linewidth]{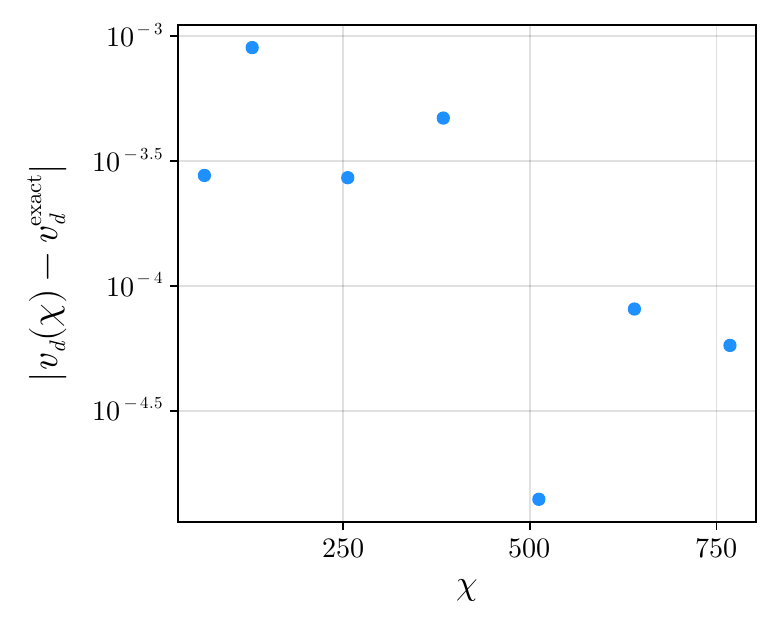}}
        \caption{(Left). Drift velocity as a function of $\mu$, for Haar-random charge-conserving gates. The solid line is the exact expression, the dots are instead extracted from the numerical simulations. The error bars show the difference between results obtained with bond dimensions $\chi=256$ and $512$. Both plots refer to $s_1 = 1$, $s_2 = \frac 1 2$. (Right). Convergence of the drift velocity $v_d(\chi)$ extracted from the numerical simulations to the exact value $v_d^\text{exact} = 5/11$. Data refer to $\mu=0$ and gates randomly drawn according to the Haar measure, with the constraint of charge conservation, and considering a single realization. The absolute error is shown as a function of the bond dimension $\chi$. }
        \label{fig:plots-convergence-and-multi-mu}
    \end{figure}

    In \figref{fig:plots-convergence-and-multi-mu} we plot the numerically obtained drift velocity against the theoretically predicted values for $s_1=1$, $s_2 = \frac 1 2$.
    Furthermore, we performed the same calculations for $s_1=1$ and $s_2=\frac 3 2$ and observed results matching the predicted drift velocity. 
    Finally, for the numerically more tractable case $s_1=1$, $s_2 = \frac 1 2$, we also verified that the drift velocities have reached convergence in the bond dimension $\chi$.

    \subsection*{Numerical simulations in the translation-invariant case}
    The numerical results shown in \figref{fig:plots-haar-break} (a), for a homogeneous circuit, have been obtained by applying a fixed, charge-conserving gate throughout the entire brickwork circuit. This gate has been obtained by sampling the Haar distribution for each of the magnetization sectors, and its non-zero entries are listed in the following:
    \begin{align*}
        \bra{+1,\uparrow} ~ A ~ \ket{+1,\uparrow} = & ~ 0.658116 + 0.752917 ~ i \\
        \bra{+1,\downarrow} ~ A ~ \ket{+1,\downarrow} = & ~ -0.127428 + 0.08373 ~ i \\
        \bra{0,\uparrow} ~ A ~ \ket{+1,\downarrow} = & ~ 0.946525 - 0.284328 ~ i \\
        \bra{+1,\downarrow} ~ A ~ \ket{0,\uparrow} = & ~ -0.615576 + 0.773186 ~ i \\
        \bra{0,\uparrow} ~ A ~ \ket{0,\uparrow} = & ~ -0.125073 + 0.0872088 ~ i \\
        \bra{0,\downarrow} ~ A ~ \ket{0,\downarrow} = & ~ 0.448503-0.626746 ~ i \\
        \bra{-1,\uparrow} ~ A ~ \ket{0,\downarrow} = & ~ -0.504415-0.389358 ~ i \\
        \bra{0,\downarrow} ~ A ~ \ket{-1,\uparrow} = & ~ -0.448171+0.452965 ~ i \\
        \bra{-1,\uparrow} ~ A ~ \ket{-1,\uparrow} = & ~ -0.527849-0.561552 ~ i \\
        \bra{-1,\downarrow} ~ A ~ \ket{-1,\downarrow} = & ~ 0.895024 - 0.446019 ~ i .
    \end{align*}
    
    \subsection*{Details on the alternating patterns breaking translation invariance}
\begin{figure}[h]
        \centering
        \includegraphics[width=0.85\linewidth]{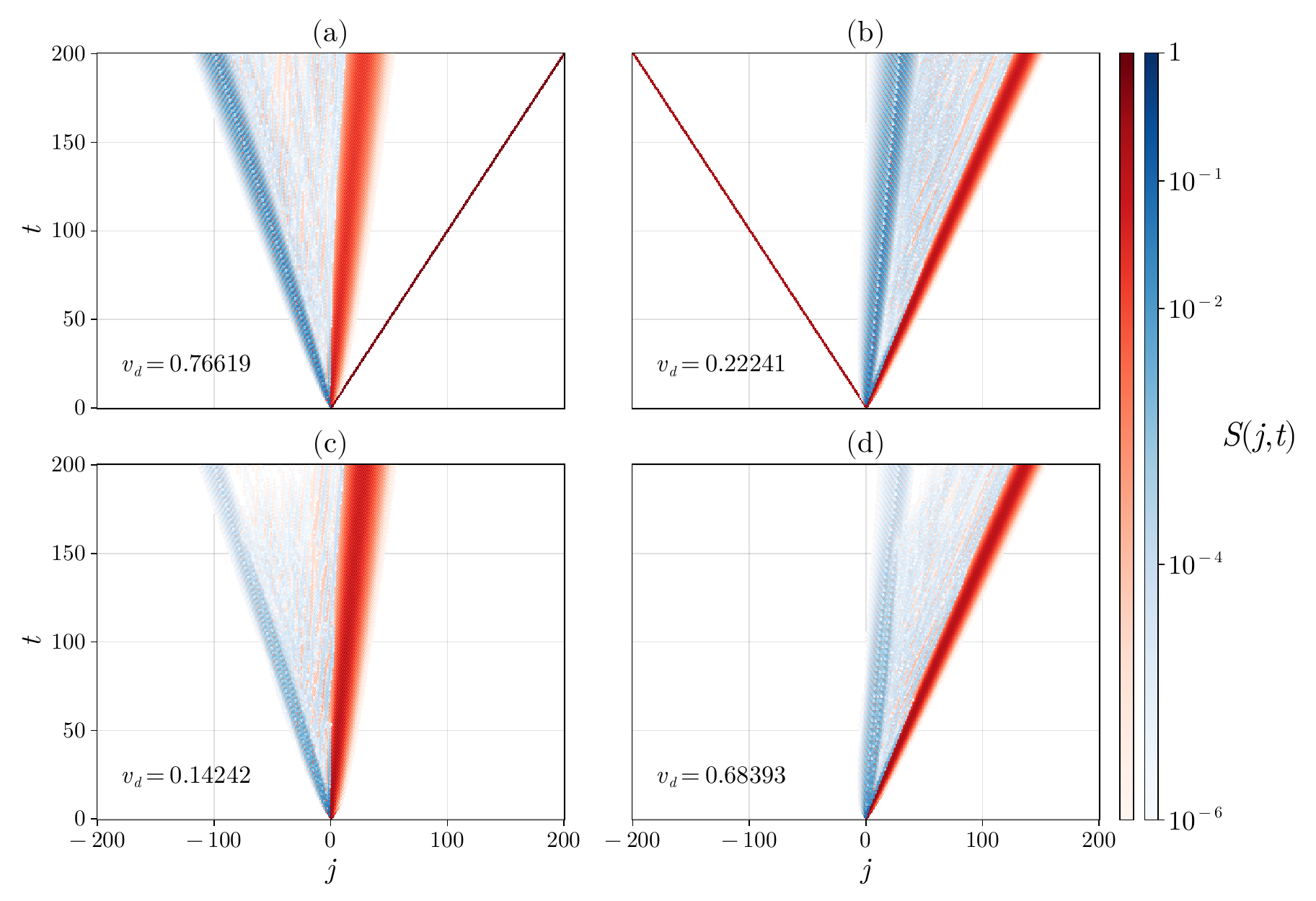}
        \caption{Two-point functions for different alternating patterns with $m=2$, as detailed in \figref{fig:alternating-pattern-schematic} for each panel. The heatmaps show positive (negative) values in red (blue); the drift velocity is annotated on each plot. The simulations refer to $\mu=0$, $s_1 = 1$, $s_2 = \frac 1 2$. Panel (c) corresponds to panel (c) of \figref{fig:plots-haar-break} in the main text. The straight lines in panels (a) and (b) show that an operator is propagating ballistically, not being acted upon by $G$ gates. Averaging $v_d$ between (a) and (c), and analogously between (b) and (d), yields the result $v_d\approx 5/11$ as expected in the homogeneous circuit.}
        \label{fig:plots-mod}
    \end{figure}

\begin{figure}[h]
        \centering
        \includegraphics[width=0.55\linewidth]{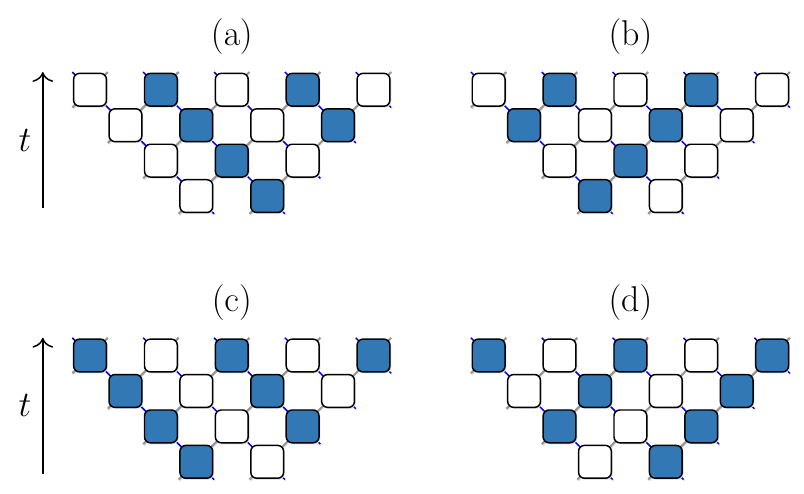}
        \caption{Different alternating patterns with $m=2$, matching those used to generate \figref{fig:plots-mod}. Each circuit refers to the corresponding panel of \figref{fig:plots-mod}. For each layer, only gates inside the light-cone are shown; gates for which $V=\mathbb I$ are depicted as white, while blue is used for $G$.}
        \label{fig:alternating-pattern-schematic}
    \end{figure}

\begin{figure}[h]
        \centering
        \includegraphics[width=0.6\linewidth]{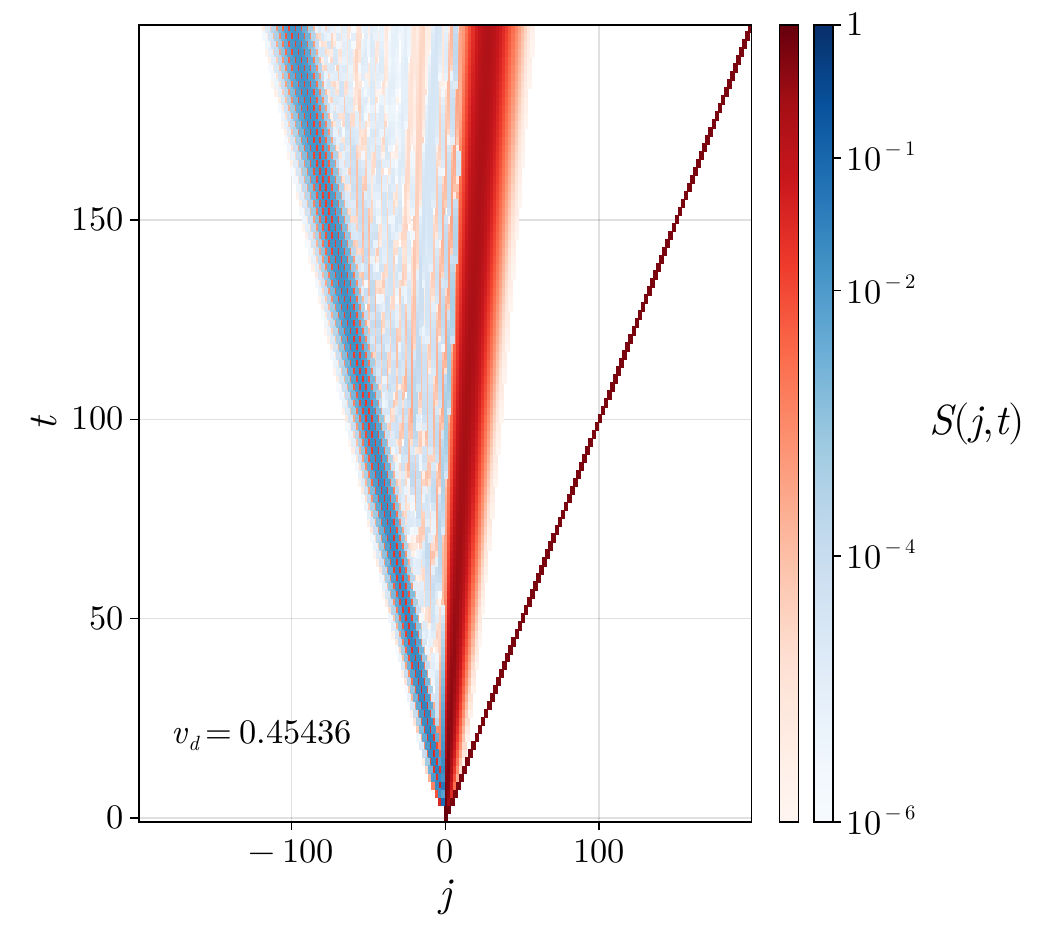}
        \caption{Two-point function for the 4-site magnetization. The ratchet used here is the same as the one in panel (c) of the previous Figs.~\ref{fig:plots-mod} and \ref{fig:alternating-pattern-schematic}. The drift velocity is the expected drift velocity in the homogeneous case.}
        \label{fig:plot-4sites-blocked}
    \end{figure}

    To obtain different drift velocities, we consider the alternating patterns with the identity and the unitary $G$ described in the main text. Here, we show the two-point function (see \figref{fig:plots-mod}) in the case $m = 2$ for several definitions of such a pattern (\figref{fig:alternating-pattern-schematic}). We observe several drift velocities, all different from the one expected in the homogeneous case. However, translation invariance is restored by considering instead a coarse-grained operator $\mathcal Z_{4n} = \sum_{\ell = 0}^3 S^z_{4n+\ell}$, for which we numerically retrieve the drift velocity expected in the homogeneous case $v_d = 5/11$ (see \figref{fig:plot-4sites-blocked}).
    
\end{document}